\documentclass[reprint,amsmath,amssymb,aps,prd]{revtex4-2}
\usepackage{graphicx}
\usepackage{dcolumn}
\usepackage{bm}
\usepackage{amsmath}
\usepackage{hyperref}
\hypersetup{
colorlinks=true,
linkcolor=blue,
citecolor=red}

\begin{document}
\title{Near Equilibrium Constraints on Bulk Viscous Models in $f(R,T)=R+2\lambda T$ Gravity}
 \author{Vishnu A Pai}
 \email{vishnuajithj@gmail.com, vishnuajithj@cusat.ac.in}
\author{Titus K Mathew}
 \email{titus@cusat.ac.in}
\affiliation{Deparment of Physics, Cochin University of Science and Technology, Cochin
}
\date{\today}

\begin{abstract}
	
	Recent studies indicate that, near equilibrium condition could not be maintained for bulk viscous matter models during the accelerated expansion of the universe in the context of Einstein's gravity, without including the cosmological constant. But from our investigation in $f(R,T)$ gravity, it is observed that, this condition can be satisfied in this modified gravity regime by properly constraining the coupling and viscous parameters. Accordingly, strict constraints are developed for free parameters in bulk viscous models in $f(R,T)=R+2\lambda T$ gravity based on fulfillment near equilibrium condition. 
	Then, for assessing the validity of NEC during different stages of evolution, two cosmological models are studied for each case based on the developed constraints. Initially, the data analysis of the models is performed using the Observational Hubble Data (OHD) and then later, model showing the best result is analyzed using combined OHD+SNe Ia data sets. From the obtained best fit values of model parameters, inferences are made regarding the possibilities of achieving recent acceleration for viscous models in $R+2\lambda T$ gravity while simultaneously satisfying the required conditions both in the presence and absence of cosmological constant.
\begin{description}
\item[Keywords]
Bulk viscosity, f(R,T) gravity, Near equilibrium condition. 
\end{description}
\end{abstract}
\maketitle

\section{Introduction}

First observational evidence supporting an accelerated expansion of the universe was obtained form the SNe Ia data by Riess et al. in 1998 and Perlmutter et al. in 1999 \cite{SupernovaSearchTeam:1998fmf,1999ApJ...517..565P}. Since then, several observations have been made in support of this celebrated discovery \cite{Bennett_2003, PhysRevD.69.103501, LIGOScientific:2017adf, Riess:2021jrx}. In light of which, an exotic cosmic component, named dark energy was proposed for explaining this recent acceleration.  Plenty of models have been put forward to model this dark energy and hence to explain the possible cause of this acceleration. Among them, the standard $\Lambda$CDM model, is accepted as the concordance model, which considers the cosmological constant $\Lambda$ as the dark energy component of the universe.  However, the model suffers with severe problems like cosmological constant problem \cite{ RevModPhys.61.1, doi:10.1146/annurev.aa.30.090192.002435, Carroll:2000fy, PADMANABHAN2003235} and coincidence problems \cite{PhysRevLett.82.896, Steinhardt:1999nw}. This lead to the introduction of several varying dark energy models \cite{doi:10.1142/S021827180600942X, PhysRevD.26.2580, PhysRevLett.81.3067, PhysRevD.62.023511, Kamenshchik:2001cp}, where the dark energy density dilutes with the expansion of the universe. In addition, as another way of explaining this late acceleration, various modified gravity theories such as $f(R)$ gravity \cite{doi:10.1142/S0218271802002025}, Lovelock gravity \cite{PADMANABHAN2013115}, $f(T)$ gravity \cite{PhysRevD.75.084031}, Horava–Lifshitz
gravity \cite{PhysRevD.79.084008}, scalar–tensor theories \cite{PhysRevD.60.043501},  Gauss Bonnet theory \cite{nojiri2005gauss}, braneworld models \cite{DVALI2000208} etc., were also proposed, which try to mimic the dark energy component in many ways.  In spite of all these fruitful efforts, the origin, nature and evolution of the dark energy component still remains a mystery.  

As a possible and natural solution to these issues, models were proposed that could explain the late accelerated expansion of the universe, without assuming any exotic dark energy component. The prominent among them are a class of  models which invokes bulk viscosity associated with the matter sector of the universe, for achieving the necessary negative pressure to explain this late acceleration. It is quite obvious to assume that, the concept of an ideal fluid as used in the concordance model could be an approximation to reality and hence, dissipative characteristics of the fluid is to be taken into account in order to gain a better understanding about the evolution of the universe, including late acceleration. From previous studies in inflationary cosmology \cite{RMaartens_1995, PhysRevD.33.1839}, it was already established (well before the SNe Ia observations) that, fluids with bulk viscosity has the innate ability to predict the accelerated expansion of the universe, that too in the absence of cosmological constant. This fact motivate several authors to investigate the possibilities of achieving recent acceleration of the universe as caused by  viscous fluid alone without incorporating the cosmological constant or dark energy. 

Most widely investigated theory in viscous cosmology is based on Eckart's theory \cite{PhysRev.58.919}. Later, an equivalent formalism was proposed also by Landau and Lifshitz \cite{landau2013fluid}. Despite of having severe drawbacks such as possessing an acausal nature or having unstable equilibrium states, the above  theory proves to be a good first order approximation for investigating viscous nature in cosmological context. In more general case, second order theories such as Muller Israel Stewart theory (MIS) or its Truncated versions \cite{1967ZPhy..198..329M, Israel:1976tn, ISRAEL1976213, Israel:1979wp}, which are associated with higher order differential equations, are also adequately discussed in literature. However, recently more general formalism for incorporating bulk viscosity with comic matter, in irreversible processes, have been proposed in references\cite{PhysRevE.56.6620, PhysRevD.60.103507, PhysRevD.61.023510, OTTINGER1998433} and also in \cite{PhysRevD.98.104064, PhysRevD.100.104020, Bemfica:2020zjp, Kovtun:2019hdm, Hoult:2020eho}. Nevertheless, since the mentioned theories pose difficulties in cosmological modelling, Eckart theory still remains as a primary choice for carrying out preliminary studies in viscous cosmology. 

Eckart's theory was developed under the condition that universe undergoes a quasi-static expansion and the concerned viscous fluid component always remains in a state of equilibrium or near to it. Hence, strictly speaking, the applicability of this theory remains confined only to those cases where the viscous fluid remains in a near-equilibrium state. This was first pointed out by Marteens in his study on dissipative inflation \cite{RMaartens_1995}. Alternatively the author points out that, one can still apply this theory in regimes where fluid is far-from equilibrium, provided one should be willing to postulate that this theory still holds in that domain. Following this assumption, several models have been proposed to explain recent acceleration of the universe
 \cite{mohan2017bulk, sasidharan2015bulk, avelino2009can, avelino2010exploring, brevik2017viscous}, which are reasonably successful in predicting cosmological parameters within an acceptable range. However the safe side is to assume that the near equilibrium condition (NEC) for the viscous fluid should be satisfied. In mathematical terms this condition suggests that the magnitude of viscous pressure of the fluid should not exceed the magnitude of its kinetic pressure, i.e.,
 \begin{equation}
	\bigg | \frac{\Pi}{p_{0}}\bigg | <<1.
\end{equation}
 In recent studies by Cruz et al.\cite{Cruz:2022zxe,sym14091866,PhysRevD.105.024047}, it was inferred that the near equilibrium condition for the viscous matter could in fact be obeyed during the accelerated expansion of the universe, but only in the presence of a cosmological constant. In addition, the authors also ruled out the possibility of assuming bulk viscosity to cold dark matter (CDM) based on the fact that the kinetic pressure of CDM is zero and hence NEC is always violated. These works are done assuming the Einstein's gravity as the background.
 Consequently, it is of great significance to investigate the possibilities of achieving recent acceleration of the universe by viscous models, satisfying the NEC for the viscous fluid, possibly without the inclusion of cosmological constant, in a modified gravity realm. Modified gravity theories are of much importance for obtaining suitable description of gravity at large scales, relevant to study of the universe. One such method is to replace the standard Einstein-Hilbert action with a suitable function of the Ricci scalar $(R)$. These types of modified gravity theories are known as $f(R)$ models \cite{capozziello2003curvature, allemandi2004accelerated, nojiri2004modified, dolgov2003can, nojiri2006modified, carroll2004cosmic}, and are extensively used in the context of cosmology. For a detailed review on this topic refer  \cite{nojiri2011unified, sotiriou2010f, de2010f}. An extension of $f(R)$ models, called the $f(R,T)$ models of gravity has also been proposed \cite{harko2011f}, in which the gravitation Lagrangian is an arbitrary function of $R$, the Ricci scale scalar and  $T$, the trace of energy momentum tensor. The dependence on $T$ in this theory may be assumed to be induced by some imperfect fluids or through quantum effects. 
One of the interesting features of this modified gravity theory is that, the particles possess an additional acceleration factor due their non-geodesic motion arising out of non-zero covariant divergence of energy momentum tensor. In the present paper we consider $f(R,T)$ gravity, with bulk viscous fluid, to explain the late acceleration, by satisfying NEC.
 
Considering bulk viscous fluids in $f(R,T)$ models is not a new concept, but all the previous studies were made without caring the NEC \cite{Koussour:2021flk, baffou2015cosmological, Prasad:2020qug, singh2014friedmann, Arora:2020xbn,Debnath:2020bno,Yadav:2020vih}.  Therefore, studies on these models by taking care of NEC is of utmost importance as the basic theories of bulk viscosity, for instance Ekart theory, are proposed by assuming NEC. Hence our primary aim is to check, whether it is possible to achieve the recent acceleration of the universe with bulk viscous fluid in these types of models, by satisfying the NEC. In the present work, we effectively study two distinct cosmological models of f(R,T) with bulk viscous fluid. The first model is one, which has a cosmological constant, and the second one is devoid of it.
 
In our analysis, we will first obtain the general constraints on the model parameters in both models using NEC and the critical energy condition, $\rho>0.$ The constraints will be obtained by assuming the equation of state associated with kinetic pressure of the cosmic fluid to be a constant negative or positive value, except zero. Then we will derive the general solution for the Hubble parameter, by assuming a special case of the parameterized form of coefficient of bulk viscosity given in \cite{Gomez:2022qcu} as, $\zeta=\zeta_0 H^{1-2s}\rho^{s}$. Here, $\zeta_0$ is a constant, $\rho$ is the energy density of the fluid and $H$ is the Hubble parameter of the universe. We have considered $s=1$ case, for which the viscous coefficient takes the form $\zeta=\zeta_0 H^{-1}\rho$. We  will study the cosmic evolution both in the presence and absence of a positive cosmological constant. Then, we compare each case with the observational data, subjected to constraints developed and thus extract the model parameters. 
	
\section{ $f(R,T)$ gravity with generic bulk viscous fluid }
$f(R,T)$ gravity is a  generalization of $f(R)$ gravity, where, the Einstein-Hilbert action is modified 
such that, the gravitational Lagrangian is treated as an arbitrary function of $R,$ the Ricci scalar and $T,$ the trace of energy momentum tensor associated with matter Lagrangian $L_{m}$. Then the modified action integral takes the form \cite{harko2011f},
\begin{equation}\label{1}
	S = \frac{1}{16\pi}\int f(R,T)\sqrt{-g}d^{4}x + \int L_{m}\sqrt{-g}d^{4}x
\end{equation}
Particular classes of $f(R,T)$  gravity models can be obtained by suitably choosing the functional form of $f(R,T)$. A plausible choice for this function, which is often used in literature is $f(R,T)=R+2f(T),$ where $f(T)$ is an arbitrary function of the trace of the matter stress-energy tensor. The field equations corresponding to this form of $f(R,T)$ is obtained in \cite{harko2011f}. Accordingly, a simple cosmological model arises if one chooses $f(T)=\lambda T$\cite{harko2011f}.  

Here, we plan on investigating two models for the universe, i.e.. one in which cosmological constant is absent and other in which it is present, hence, we are interested in two particular form of $f(R,T)$ given by $f(R,T)= R + 2\lambda T$ and $f(R,T)= \Lambda + R + 2\lambda T$ respectively. Here $\lambda$ is having the status of a coupling constant that explicitly connects the energy content of the universe to gravitational Lagrangian $L_{g}$. We consider a single component universe, comprising of a generic bulk viscous fluid with matter Lagrangian $L_{m}$ and associated energy momentum tensor $T_{\mu\nu}$ having its trace $T$ coupled to gravity as shown by equation (\ref{1}). For the first case, extremizing the action in equation (\ref{1}) by assuming $f(R,T)=R+2\lambda T,$ we get the modified field equation as,
\begin{equation}\label{2}
	R_{\mu\nu}-\frac{1}{2}Rg_{\mu\nu}= 8\pi T_{\mu\nu} -2\lambda\left[ T_{\mu\nu}+\Theta_{\mu\nu}-Tg_{\mu\nu}\right],
\end{equation}
where, $\Theta_{\mu\nu}$ is defined as 
\cite{harko2011f},
\begin{equation}\label{3}
	\Theta_{\mu\nu}=-2T_{\mu\nu}+g_{\mu\nu}L_{m}-2g^{\alpha_{1}\beta_{1}}\frac{\partial^{2}L_{m}}{\partial g^{\mu\nu}\partial g^{\alpha_{1}\beta_{1}}},
\end{equation}
in which $\alpha_1$ and $\beta_1$ are indices to be summed over.
The filed equation can equivalently be expressed as,
\begin{equation}\label{4}
	R_{\mu\nu}-\frac{1}{2}Rg_{\mu\nu}= 8\pi T^{eff}_{\mu\nu}
\end{equation}
where,
\begin{equation}\label{5}
    T^{eff}_{\mu\nu} =  T_{\mu\nu} - \frac{2\lambda\left[ T_{\mu\nu}+\Theta_{\mu\nu}-Tg_{\mu\nu}\right]}{8\pi}
\end{equation}
is an effective energy-momentum tensor.

Now, for the second case, the extremization of the action (\ref{1}) by assuming $f(R,T)=\Lambda + R+2\lambda T,$ leads to the field equation,
\begin{equation}\label{1.2}
R_{\mu\nu}-\frac{1}{2}Rg_{\mu\nu}+\Lambda g_{\mu\nu}= 8\pi T_{\mu\nu} -2\lambda\left[ T_{\mu\nu}+\Theta_{\mu\nu}-Tg_{\mu\nu}\right],
\end{equation}
Where, $\Theta_{\mu\nu}$ has same definitions as equation (\ref{3}). This also can be written in an equivalent form by defining an effective energy-momentum tensor as in equation (\ref{5}), as, 

\begin{equation}\label{1.4}
R_{\mu\nu}-\frac{1}{2}Rg_{\mu\nu}+\Lambda g_{\mu\nu}= 8\pi T^{eff}_{\mu\nu}
\end{equation}
In both the above cases, the stress-energy momentum tensor of bulk viscous fluid is taken in conventional form as,
\begin{equation}\label{6}
	T_{\mu \nu} = (\rho+ p) u_{\mu}u_{\nu}+pg_{\mu \nu}
\end{equation}
Where, $p=p_{0}+\Pi$ with $p_{0}=\omega \rho.$ Here $p_{0}$ corresponds to the kinetic pressure of the viscous fluid and $\Pi$ corresponds to its bulk viscous pressure. For maintaining a general nature for the constraints that are developed, we consider a constant equation of state parameter ($\omega$) which can take both negative and positive values but not zero. We have omitted the case where $\omega =0$ because, in that scenario, viscous fluid has no kinetic pressure and hence required condition (NEC) that we are trying to impose will always get violated. The viscous coefficient is also assumed to take both negative or positive values. The dynamics of the universe in these two models can then be studied based on the assumption that the space is flat, homogeneous and isotropic, hence described by the FLRW metric having line element,
\begin{equation}\label{7}
	ds^{2}=g_{\mu\nu}dx^{\mu}dx^{\nu} = -dt^2+a(t)^2\left(dx^{2}+dy^{2}+dz^{2}\right).
\end{equation}
Friedmann equations for the first case is then obtained, by substituting equations (\ref{6}) and (\ref{7}) in the field equation (\ref{4}), as,
\begin{equation}\label{8}
	3H^{2}=\rho^{eff}=\rho+\tilde{\lambda} \left( 3\rho-p\right)
\end{equation}
\begin{equation}\label{9}
	2\dot{H}+3H^{2}=-p^{eff}=-\left[\; p +\tilde{\lambda} \left( 3p-\rho \right)\;\right].
\end{equation}
Here an over-dot represents a derivative with respect to the cosmic time, $t.$
Similarly, Friedmann equations for the second case can be obtained by using equations (\ref{6}) and (\ref{7}) in the field equation (\ref{1.4}), and is,
\begin{equation}\label{1.8}
3H^{2}=\rho+\tilde{\lambda} \left( 3\rho-p\right)+\Lambda
\end{equation}
\begin{equation}\label{1.9}
2\dot{H}+3H^{2}=-\left[\; p +\tilde{\lambda} \left( 3p-\rho \right)-\Lambda\;\right]
\end{equation}
where,
\begin{equation}\label{10}
	p=p_{0}+\Pi=\omega \rho +\Pi
\end{equation}
For convenience, we have re-scaled  $\lambda$ to $\tilde{\lambda}=\lambda c^4/ 8 \pi G$ and have chosen $8 \pi G/c^4=1$. The continuity equation for the fluid can be obtained by combining equation of state (EoS) of fluid $p_{0}=\omega \rho$ with the respective Friedmann equations, and it takes a general form, true for both cases, as,
\begin{equation}\label{11}
	\dot{\rho}+3H \left( \rho+p \right)\left(\frac{1+2\tilde{\lambda}}{1+3\tilde{\lambda}}\right)=\dot{p}\left(\frac{\tilde{\lambda}}{1+3\tilde{\lambda}}\right).
\end{equation}
Hence, as expected, the conservation equation for the fluid in $f(R,T)$ gravity differs from that obtained from Einstein-Hilbert action. Note that all the above equations reduce back to the corresponding equations in Einstein's gravity, if one sets $\tilde{\lambda}=0$.

\section{NEC and the Constraints on the parameters. } \label{section 3}

 Eckart's theory for a relativistic dissipative fluid was developed under the assumption that, throughout the evolution, a viscous fluid always remains in a state of equilibrium or near to it.  In other words, magnitude of the bulk viscous pressure must be smaller than kinetic pressure of the fluid i.e,
 \begin{equation}\label{12}
 	\bigg | \frac{\Pi}{p_{0}}\bigg | <<1.
 \end{equation}
Hence any deviation from this condition suggest that fluid is far from equilibrium and causes the theory to breakdown. This was first pointed out by Marteens \cite{RMaartens_1995} in his study on dissipative cosmology, in the context of early inflation. However there exists studies in literature, in which one may blindly postulate the validity of Eckart theory\cite{mohan2017bulk, sasidharan2015bulk, avelino2009can, avelino2010exploring, brevik2017viscous, Koussour:2021flk, baffou2015cosmological, Prasad:2020qug, singh2014friedmann, Arora:2020xbn,Debnath:2020bno,Yadav:2020vih} even in such unlikely scenario where the fluid is far from equilibrium, and hence analyze the possibilities of viscous generated accelerated expansion. 
 
 From (\ref{12}) it is evident that, the NEC cannot be satisfied in models with viscous cold bulk matter (vCDM), since its kinetic pressure is zero. One may then check the possibilities of generating accelerated expansion by satisfying NEC, using viscous warm dark matter (vWDM) as the cosmic component, since it can have a non-zero kinetic pressure. But in some of the recent works by Cruz et al.\cite{Cruz:2022zxe,sym14091866,PhysRevD.105.024047}, in the context of Einstein's gravity, it was argued that NEC is still violated, even when one use vWDM to describe the recent acceleration of the universe. However, positively, they found that, the validity of NEC can be regained, in such models, by incorporating a positive cosmological constant. It is then reasonable to check, whether it is possible to explain the late acceleration caused by vWDM, by satisfying NEC without a cosmological constant, in the context of alternative theories of gravity. We are analysing this possibility in $f(R,T)$ gravity using a generic bulk viscous, instead of mere vWDM, as cosmic component. 
 
 We assume that the universe is dominated by a generic bulk viscous fluid which can have a constant EoS parameter $\omega $ which may be positive or negative but excluding zero. Before analyzing the dynamics of the evolution, we will first obtain the constraints for free parameters, imposed by the NEC.  For finding the constraints, we consider two cases, one without a cosmological constant and the other is with a positive cosmological constant. At this stage it is important to note that, the constraints that we aim on developing are general in the sense that we will not be assuming any functional form for coefficient of bulk viscosity. Hence, while carrying out data analysis of a model obtained by assuming some form for coefficient of bulk viscosity, the general constraints remains unchanged while the NEC condition (\ref{12}) needs to be properly accounted based on the assumed functional form of $\zeta$. Also, these constraints are applicable only in $R+2\lambda T$ gravity.

\subsection{Constraints in the absence of cosmological constant} \label{section 3 a}

Let us first rewrite the the NEC given in (\ref{12}) in a more convenient form, which we need for later analysis, as,
  \begin{equation}\label{13}
 	-1<< \frac{\Pi}{p_{0}} <<1
 \end{equation}
By combining the two Friedmann equations (\ref{8}), (\ref{9}) with EoS (\ref{10}) we can determine the acceleration equation as,
\begin{equation}\label{14}
	\frac{\ddot{a}}{a} = -\frac{1}{6}\left( (3+8\tilde{\lambda})(p_{0}+\Pi) +\rho \right).
\end{equation}
For ensuring an accelerated expansion we require the condition, $\ddot{a}/{a}>0.$ Impossing this on the above equation leads to the condition,
\begin{equation}\label{16}
	(3+8\tilde{\lambda})(p_{0}+\Pi) +\rho <0.
\end{equation}
Here we consider $\rho>0$ always. The exact nature of the above inequality is depends on the sign of $(3+8\tilde{\lambda})$ and also on the nature of the kinetic pressure of the fluid, $p_0.$ Following this we can have different cases: 
\subsubsection{\textbf{For $(3+8\tilde{\lambda})<0$ (or $\tilde{\lambda}<-3/8$)}}
The above inequality takes the form, 
\begin{equation}\label{17}
	-\Pi <p_{0} + \frac{\rho}{3+8\tilde{\lambda}}.
\end{equation}
This very condition can be bifurcated by considering two different nature for the equation of state of the fluid component, i.e. $p_{0}=\omega \rho$. Here, $p_{0}>0$ corresponds to vWDM like behavior and $p_{0}<0$ corresponds to dark fluid like behavior.
\begin{itemize}
	\item For $\omega>0$ (that is $p_{0}>0$), we get,
	\begin{equation}\label{18}
		-\frac{\Pi}{p_{0}} < 1 + \frac{1}{\omega(3+8\tilde{\lambda})}
	\end{equation}  
	Left hand side of this inequality is a positive number if  $\Pi<0$ and $p_0>0$. This will then satisfy the NEC if we constraint $\omega$ and $\tilde{\lambda}$ using $-1<1/\omega(3+8\tilde{\lambda})<0.$ Another interesting possibility   
	for the non-violation of NEC, even with $\Pi>0,$ is 
	\begin{equation}\label{18.1}
		\frac{\Pi}{p_{0}} > -\left(1 + \frac{1}{\omega(3+8\tilde{\lambda})}\right)
	\end{equation}
	with $-2<1/\omega(3+8\tilde{\lambda})<0. $ In this case, if  $\epsilon = -\left(1 + \frac{1}{\omega(3+8\tilde{\lambda})}\right)$ is the range $\epsilon \in (-1,1)$, then NEC is not necessarily
	\item  $\omega<0$ (or $p_{0}<0$):
\begin{equation}\label{20}
	-\frac{\Pi}{p_{0}} > 1 + \frac{1}{\omega(3+8\tilde{\lambda})}
\end{equation}  
In this case, since $3+8\tilde{\lambda}<0$ and $\omega<0$, the second term on right side is positive which implies that the right side of the inequality always has a value greater than 1 and therefore (\ref{20}) violates NEC.
\end{itemize}
The above analysis shows that, for $\tilde{\lambda}<-3/8$, NEC is satisfied only when the generic viscous fluid show vWDM like traits with $\omega>0$. 
\subsubsection{ \textbf{For $(3+8\tilde{\lambda})>0$ (or $\tilde{\lambda}>-3/8$)}}
Similar to earlier case, we rearrange the inequality (\ref{16}) and divide either side by $(3+8\tilde{\lambda})$. Then, the inequality doesn't flip direction since $(3+8\tilde{\lambda})>0,$ and we have, 
\begin{equation}\label{21}
-\Pi > p_{0} + \frac{\rho}{3+8\tilde{\lambda}}
\end{equation}
Both sides of inequality (\ref{21}) is divided by $p_{0}=\omega \rho$, then we get two sub cases similar to those found in the above case.
\begin{itemize}
\item  $\omega>0$ (that is $p_{0}>0$), we get,\\
\begin{equation}\label{22}
	-\frac{\Pi}{p_{0}} > 1 + \frac{1}{\omega(3+8\tilde{\lambda})}
\end{equation}  
Since, we have $\omega>0$ and $3+8\tilde{\lambda}>0$, the second term on right side is always positive which makes the right hand side of the above inequality is always greater than one. Hence, the NEC is always violated in this scenario.
\item $\omega<0$ (or $p_{0}<0$ ):\\
\begin{equation}\label{23}
	-\frac{\Pi}{p_{0}} < 1 + \frac{1}{\omega(3+8\tilde{\lambda})}
\end{equation} 
Here we have, $\omega<0$ and $3+8\tilde{\lambda}>0$, which makes the value on right side less than one. To make sure that the NEC remains satisfied, we must further constrain values of $\omega$ and $\tilde{\lambda}$ as,
\begin{equation}\label{24}
	\begin{cases}
		-1<\frac{1}{\omega(3+8\tilde{\lambda})}<0,& \text{if } \Pi >0\\
		-2<\frac{1}{\omega(3+8\tilde{\lambda})}<0, & \text{if } \Pi <0\\
	\end{cases}
\end{equation}  

\end{itemize}
Hence, for $\tilde{\lambda}>-3/8$, the NEC is satisfied only when generic fluid shows behavior of viscous dark energy like fluid with $\omega <0$ and not when it show vWDM behavior with $\omega >0$. 
To sum up, there are definite constraints, as discussed above, on $\tilde{\lambda}$ and $\omega$, if one intends to explain recent acceleration while simultaneously satisfying NEC for the viscous fluid. For given model, the status of these constraints during different phases of evolution can depend on how Hubble parameter changes in time and thus depends on evolutionary behaviors of $\rho$ and $\Pi$.

\subsubsection{\textbf{Constraints based on Critical Energy Condition }}
In this part, we check, in detail the additional constraints emerging form the requirement of NEC constraint, i.e. $\rho>0$ or equivalently $\Omega_{\rho}>0. $ This constraint on $\Omega_{\rho}$ is what we call as the critical energy condition. It can be seen that this condition differ from weak energy condition $\rho_{eff}\geq0$, which is proposed in case of modified gravity theories. But this new constraint is a necessary condition as far as satisfying the NEC for the viscous fluid using the previously developed constraints are concerned. Combine equations (\ref{8}) and (\ref{10})  gives,
\begin{equation}\label{26}
3H^{2}=\{(1+\tilde{\lambda}(3-\omega))\rho\} -\tilde{\lambda}\Pi
\end{equation}
On rearranging the above equation we get,
\begin{equation}\label{27}
\Omega_{\rho}=\frac{\rho}{3H^{2}}=\frac{1+\tilde{\lambda}\Omega_{\Pi}}{1+\tilde{\lambda}(3-\omega)}
\end{equation}
Where, $\Omega_{\Pi}=\Pi/3H^{2}.$ Then $\Omega_{\rho}>0$ implies,
\begin{equation}\label{28}
\frac{1+\tilde{\lambda}\Omega_{\Pi}}{1+\tilde{\lambda}(3-\omega)}>0
\end{equation}
This condition can hold for the following two distinct cases.
\begin{itemize}
\item For $1+\tilde{\lambda}\Omega_{\Pi}>0$ and $1+\tilde{\lambda}(3-\omega)>0$ 
\begin{enumerate}
	\item if  $\tilde{\lambda}>0$, we get the constraints on $\Omega_{\Pi}$ and $\omega$ as,
	\begin{equation}\label{29}
		\Omega_{\Pi}>\frac{-1}{\tilde{\lambda}} \text{ \; and \; } 	(3-\omega)>\frac{-1}{\tilde{\lambda}}
	\end{equation}
	
	\item if $\tilde{\lambda}<0$ ,we get the corresponding constraints as,
	\begin{equation}\label{30}
		\Omega_{\Pi}<\frac{-1}{\tilde{\lambda}} \text{ \; and \; } 	(3-\omega)<\frac{-1}{\tilde{\lambda}}
	\end{equation}
\end{enumerate}
\item For $1+\tilde{\lambda}\Omega_{\Pi}<0$ and $1+\tilde{\lambda}(3-\omega)<0$
\begin{enumerate}
	\item if $\tilde{\lambda}>0$, we get the constraints on values of $\Omega_{\Pi}$ and $\omega$ as,
	\begin{equation}\label{31}
		\Omega_{\Pi}<\frac{-1}{\tilde{\lambda}} \text{ \; and \; } 	(3-\omega)<\frac{-1}{\tilde{\lambda}}
	\end{equation}
	\item if $\tilde{\lambda}<0$, we get, 
	\begin{equation}\label{32}
		\Omega_{\Pi}>\frac{-1}{\tilde{\lambda}} \text{ \; and \; } 	(3-\omega)>\frac{-1}{\tilde{\lambda}}
	\end{equation}
\end{enumerate}
\end{itemize}

\subsection{Constraints in the presence of cosmological constant}\label{section 3 b}

In the previous section we have obtained the constraints on the parameters without considering the cosmological constant. But it is always worth obtaining the constraints in the presence a cosmological constant, $\Lambda.$  Hence in this section we investigate NEC in the presence of a positive cosmological constant. Since, inclusion of $\Lambda$ will anyway guarantee the presence of a cosmic component having negative equation of state, here we restrict the equation of state of the generic fluid, to be in the range, $\omega>0$ and also consider a single case where $\tilde{\lambda} >0$. By combining the modified Einstein equations (\ref{1.8}), (\ref{1.9}) and EoS (\ref{10}) and we can obtain the acceleration equation as,
\begin{equation}\label{37}
	\frac{\ddot{a}}{a} = -\frac{1}{6}\left( (3+8\tilde{\lambda})(p_{0}+\Pi) +\rho -2 \Lambda \right)
\end{equation}
For accelerated expansion, $\ddot{a}>0$, which suggests,
\begin{equation}\label{38}
	(3+8\tilde{\lambda})(p_{0}+\Pi) +\rho -2 \Lambda < 0
\end{equation}
Or equivalently we have,
\begin{equation}\label{39}
	-\frac{\Pi}{p_{0}}>1 +\frac{1}{(3+8\tilde{\lambda})\omega} -\frac{2 \Lambda}{(3+8\tilde{\lambda})p_{0}} 
\end{equation}
Here we take $\tilde{\lambda}>0$ and $\Lambda>0$. Taking account of the last negative term on the right hand side, we learn that the magnitude of $\Pi/p_{0}$ is not necessarily greater than one, and in order to make sure that it always stays that way, we impose a constraint on the value of $\Pi$ using the NEC. This is achieved by combining the condition (\ref{12}) with equations (\ref{1.8}) and EoS (\ref{10}), which then provides the necessary inequality constraint in terms of  $\Omega_{\Pi}$ and $\Omega_{\Lambda}=\Lambda/3H^{2}$ as,
 \begin{equation}\label{40}
	\Bigg | \; \frac{\Omega_{\Pi} \left(   1+\tilde{\lambda}(3-\omega)   \right)}{\omega \left(   1+\tilde{\lambda}\Omega_{\Pi} -\Omega_{\Lambda} \right)} \; \Bigg | <<1.
\end{equation}
 This single constraint developed in the presence of $\Lambda$  is powerful enough to enforce the required conditions for NEC.
 
\section{Evolution of Hubble parameter }
In this section, we obtain the analytical solution for Hubble parameter by choosing a special case of newly proposed generally parameterized form of bulk viscosity \cite{Gomez:2022qcu} which is given as $\zeta=\zeta_0 H^{1-2s}\rho^{s}$. Here, $\zeta_0$ is a constant, $\rho$ is the energy density of the fluid and $s$ is some constant number. Both in the presence and absence of the cosmological constant, we will consider $s=1$ case of this viscous model, for which viscous coefficient attains the form $\zeta=\zeta_0 H^{-1}\rho$.

\subsection{Case I: Model with $\Lambda$ }\label{case2}

Combining Friedmann equations (\ref{1.8}), (\ref{1.9}) along with relation (\ref{10}), and assuming the bulk viscous coefficient as, $\zeta=\zeta_{0}H^{-1}\rho$, a first order ODE for the Hubble parameter can be obtained as,

\begin{equation}\label{47}
	2\dot{H}  = \left[-3H^{2} + \Lambda \right] \left[\frac{(1+2\tilde{\lambda})\left((\omega+1)-3\zeta_{0}\right)}{1+\tilde{\lambda}\left(3-\omega+3\zeta_{0}\right)}\right]
\end{equation}
 To solve this equation and obtain Hubble parameter in terms of scale factor, we make a change of variables from time $t$ to scale factor $a$ using the relation, $\frac{d}{dt}=H\frac{d}{d\ln{a}}$. This changes (\ref{47}) into,
 \begin{multline}\label{48}
 	2\frac{dH}{d\ln{a}}  = \left[-3H + \frac{\Lambda}{H} \right] \left[\frac{(1+2\tilde{\lambda})\left((\omega+1)-3\zeta_{0}\right)}{1+\tilde{\lambda}\left(3-\omega+3\zeta_{0}\right)}\right]
 \end{multline}
Solving this, we get the expression for Hubble parameter as,
\begin{equation}\label{49}
		H=H_{0}\left[\Omega^0_{\Lambda} +  \left(1-\Omega^0_{\Lambda}\right)a^{-\beta_{1}}\right]^{\frac{1}{2}}
\end{equation}
Where $\Omega^0_{\Lambda}$ and $\beta_{1}$ are constants given by,
\begin{equation}\label{50}
	\Omega^0_{\Lambda}= \frac{\Lambda}{3H^2_0}
\end{equation}
\begin{equation}\label{51}
	\beta_{1}=\frac{3(2\tilde{\lambda}+1)\left((\omega+1)-3\zeta_{0}\right)}{1+\tilde{\lambda}(3-\omega+3\zeta_{0})}
\end{equation}
For $\beta_{1}>1$ and $0<\alpha_{1}<1$,  the Hubble parameter evolves such that,  (i) as $a \to 0, H \to H_0 (1-\Omega^0_{\Lambda}) a^{-\beta_1};$ and as $a \to \infty,  H \to H_0\Omega^0_{\Lambda}.$ Hence the model guarantees the transition from a prior decelerated to a later accelerated universe, which asymptotically ends on a de Sitter epoch. 

\subsection{Case II: Model without $\Lambda$ }\label{case1}

In contrast to the previous section, we will obtain the solution for the Hubble parameter without the cosmological constant. To obtain the equation for Hubble parameter in terms of scale factor, we combine the Friedmann equations (\ref{8}), (\ref{9}), (\ref{10}) and consider coefficient of bulk viscosity as $\zeta=\zeta_{0}H^{-1}\rho$ or set $\Lambda=0$ in the above case. The differential equation hence obtained for Hubble parameter is given as,
\begin{equation}\label{41}
2\dot{H}=-\left[\frac{(1+2\tilde{\lambda})\left((\omega+1)-3\zeta_{0}\right)}{1+\tilde{\lambda}\left(3-\omega+3\zeta_{0}\right)}\right]3H^{2}
\end{equation}
 Again, this differential equation can be expressed in terms of scale factor as,
\begin{equation}\label{42}
 	2\frac{dH}{d\ln{a}}  =-\left[\frac{(1+2\tilde{\lambda})\left((\omega+1)-3\zeta_{0}\right)}{1+\tilde{\lambda}\left(3-\omega+3\zeta_{0}\right)}\right]3H.
\end{equation}
Solution of this differential equation is then obtained as,
\begin{equation}\label{43}
	H=H_{0}a^{-\beta_{2}}.
\end{equation}
Where $\beta_{2}$ is a constant given by,
\begin{equation}\label{45}
	\beta_{2}=\frac{3(2\tilde{\lambda}+1)\left((\omega+1)-3\zeta_{0}\right)}{1+\tilde{\lambda}(3-\omega+3\zeta_{0})}
\end{equation}
It is clear from equation (\ref{43}) that, for $\beta_{2}<1$, this model implies an ever accelerating universe, otherwise a ever-decelerating expansion. Hence, this model is not feasible as it is not predicting the late acceleration. Nevertheless, for the purpose of checking the validity of NEC during the evolution of the universe, using the derived set of constraints, this model 
can also be used.
\begin{table*} 
	\caption{\label{tab1}Table of best estimated values of model parameters using only OHD data for Case II and using both OHD and combined OHD+SNe Ia data for Case I. }
	\begin{ruledtabular}
		\begin{tabular}{cccccccccc}
			Case&Fluid&$H_{0}$&$\tilde{\lambda}$&$\zeta_{0}$&$\Delta$&$\omega$&$\chi^{2}_{min}$&$\chi^{2}_{dof}$\\
			&Model&&&&&&
			\\ \hline
		I	& Dark Fluid&$ 66.36\pm 0.51$&$0.002 \pm0.003$&$0.01$ &$-0.07 \pm 0.02$&$-0.40\pm 0.03$&$142.22$&$2.96$\\
		I	& vWDM&$61.26 \pm 0.80$&$-0.47 \pm 0.01$& $0.03$&$0.06 \pm 0.02$&$1.41\pm 0.07$&$93.38$&$1.94$ \\
		I (using prior (\ref{prior1}))	& vWDM & $61.25 \pm 0.82$ & $0.41\pm 0.40$ &$0.23\pm 0.12$& $-$&$0.49 \pm 0.34$&$130.52$&$2.72$ 
		\end{tabular}
		\begin{tabular}{ccccccccccc}
			Case&Fluid&$H_{0}$&$\tilde{\lambda}$&$\Delta$&$\Omega^{0}_{\Lambda}$&$\zeta_{0}$&$\omega$&M&$\chi^{2}_{min}$&$\chi^{2}_{dof}$\\
			&Model&&&&&&&&&\\ \hline
II	(OHD)	&vWDM &$70.90\pm 1.39$&$0.42 \pm 0.33$&$0.45\pm0.57$&$0.72\pm 0.05$&$0.03$&$0.24\pm0.27$&$-$&$34.71$&$0.74$\\
			II  (OHD+SNe Ia)&vWDM&$ 70.17 \pm 0.73$&$0.45 \pm 0.33$&$0.49 \pm 0.52$&$0.69 \pm 0.02$ &$0.04$ &$0.26\pm0.26$&$19.35\pm0.02$&$1075.29$&$0.98$\\
		\end{tabular}
	\end{ruledtabular}
\end{table*}

\section{Estimation of model parameters subjected to NEC constraints}

Now we will compare the previous models with observational data, to extract the model parameters.  From the previous section, it becomes evident that, the model with cosmological constant is a viable one, since it predicts a transition into the late accelerated epoch.  However we will also consider the model without $\Lambda$ for academic curiosity. The data we use, for our calculations, are observational Hubble data (OHD) \cite{Geng:2018pxk} which contains 52 data points within redshift range of $0\leq z \leq 2.36$ and Type Ia Supernovae data (SNe Ia) \cite{Scolnic_2018} which contains a total of 1048 data points within a redshift range of $0.01\leq z \leq2.26$. We use the standard $\chi^{2}$ analysis method to  determine the model parameters, and using which the evolution of the observables such as deceleration parameter, transition redshift, age of the universe are to be analyzed. We will first compare the model in case II subjected to the constraints, using the OHD data. This will lead to multiple cases of same model. We will select the best among them, which will then be tested using the combined OHD and SNe Ia datasets. To carryout the $\chi^{2}$ minimization, we have employed Markov chain Monte Carlo (MCMC) estimation technique by utilizing emcee python package \cite{Foreman-Mackey_2013} in lmfit python library \cite{newville_matthew_2014_11813}. Since the constraints that where developed for free parameters are expressions in the form of inequalities, they are implemented using expression bound techniques available in lmfit library.

For our analysis using the OHD data, we compare the values obtained for theoretical Hubble parameter $H_{t}$ which is obtained for different redshifts, with those in the observational Hubble data $H_{o}$ which are also measured at different redshifts. The required $\chi^{2}$ function which is to be minimized is then given by,
\begin{equation}\label{53}
	\chi^{2}_{OHD}((a,b,...,H_{0})) = \sum_{k=1}^{n}\frac{\left[H_{t}(a,b,...,H_{0})-H_{o}\right]^{2}}{\sigma_{k}^{2}}
\end{equation}
Where, $a,b,..,H_{0}$ represents the model parameters whose best estimates are to be found, $n$ is the total number of data points available for the analysis and $\sigma_{k}^{2}$ is the variance in the measured value of  $k^{th}$ data.

To obtain the best fit value for parameters using type Ia supernovae data and to compare the best model with observational data, we use the theoretical expression for distance modulus $\mu_{t}$ of $k^{th}$ supernovae with red shift $z_{k}$ which is given by, 
\begin{align*}
	\mu_{t}(z_{k},a,b,..,H_{0})&=m-M\\
	&=5\log_{10}\left[\frac{d_{L}(z_{k},a,b,..,H_{0})}{\text{Mpc}}\right]+25 \tag{54} \label{54}
\end{align*}
Where, $m$ and $M$ are apparent and absolute magnitudes of supernovae and $d_{L}$ is the luminosity distance defined for a flat universe. Through out the analysis we treat $M$ as a nuisance parameter. The luminosity distance is given by the relation, 
\begin{equation}\label{55}
	d_{L}(z,a,b,...,H_{0})=c(1+z)\int_{0}^{z}\frac{dz'}{H(z',a,b,...,H_{0})}
\end{equation}
In this case, the required $\chi^{2}$ function is given by,
\begin{equation}\label{56}
	\chi^{2}_{SNe Ia}((a,b,...,H_{0}))  = \sum_{k=1}^{n}\frac{\left[\mu_{t}(a,b,...,H_{0})-\mu_{o}\right]^{2}}{\sigma_{k}^{2}}
\end{equation}
For the combined data analysis using OHD+SNe Ia data sets, the $\chi^{2}$ function to be minimized is then given by,
\begin{equation}\label{57}
	\chi^{2}_{total} = \chi^{2}_{OHD} +\chi^{2}_{SNe Ia} 
\end{equation}
Using these equations, a $\chi^{2}$ minimization is performed for the two models and best estimated values of the model parameters are determined. The viability of a model 
%The minimum chi-square value obtained in each case,  along with value of  chi-square degrees of freedom $\left(\chi^{2}_{dof}\right)$ for each case is also given.
is determined by the chi square per degrees of freedom, which is defined as,
\begin{equation}
	\chi^{2}_{dof}=\frac{\chi^{2}_{min}}{n-n_{p}} \notag
\end{equation}
Here, $n$ is the number of available data points and $n_{p}$ is the number of model parameters. The model is considered as, good fit to data if $\chi^{2}_{dof} \approx 1$, over fits the data if $\chi^{2}_{dof}<<1$ and bad fit to data if $\chi^{2}_{dof}>>1$.  The estimation is done by simultaneously satisfying the respective model constraints developed in section \ref{section 3 a},  \ref{section 3 b} and the requirement for having an accelerated expansion. After estimating the best fit value of model parameters, we study the evolution of cosmological important cosmological parameters, such as predicted age of the universe, deceleration parameter, transition redshift etc..

\paragraph*{\textbf{Age of the universe:}} Age of the universe predicted by the model can be determined by integrating the relation for Hubble parameter (i.e. $dt/da={[aH(a)]}^{-1}$). 
\begin{equation}
	t_{0}-t_{b}= \int_{0}^{1}\frac{da}{aH(a)} \label{57.1}
\end{equation}
Here, $t_{0}$ represents the present time (i.e at $a=1$) and $t_{b}$ is the time at which the big bang (i.e. at $a=0$) occurred. Hence $t_{0}-t_{b}$ represents the current age of the universe since the big bang.

\paragraph*{\textbf{Deceleration parameter:}} For investigating the nature of expansion of the universe or determining the period during which universe was decelerating or to find the redshift at which the transition happened from deceleration to acceleration, we require the relation for deceleration parameter. Using the scale factor, its double derivative and Hubble parameter, the deceleration parameter (q) can be defined as,
\begin{equation}
	q=-\frac{\ddot{a}}{aH^{2}}.  \label{63}
\end{equation}

\subsubsection{\textbf{With OHD data set:}}

In this section we estimate parameters using OHD data alone. For implementing the constraint given in equation (\ref{12}) for case I and case II, we  re-designate the left hand side of the same as $\Delta,$ which after substituting $\Pi$ and $p_0$ takes the same form for both cases which is given as,
\begin{equation}
\Delta=\frac{3\zeta_{0}}{\omega}.  \label{61}
\end{equation}
For case I, i.e. in the presence of $\Lambda$ we use the following priors for the relevant variables,
\begin{eqnarray} \label{prior2}
	\tilde{\lambda}\in [0,1]\;:\;\Omega^{0}_{\Lambda} \in [0,1]\;:\; \omega >0 \;:\; H_{0} \in [50,80] .
\end{eqnarray}
Here the prior for $\omega$ implies that, we consider only warm dark matter nature for the generic fluid.

\begin{figure}
	\centering
	\includegraphics[width=\columnwidth]{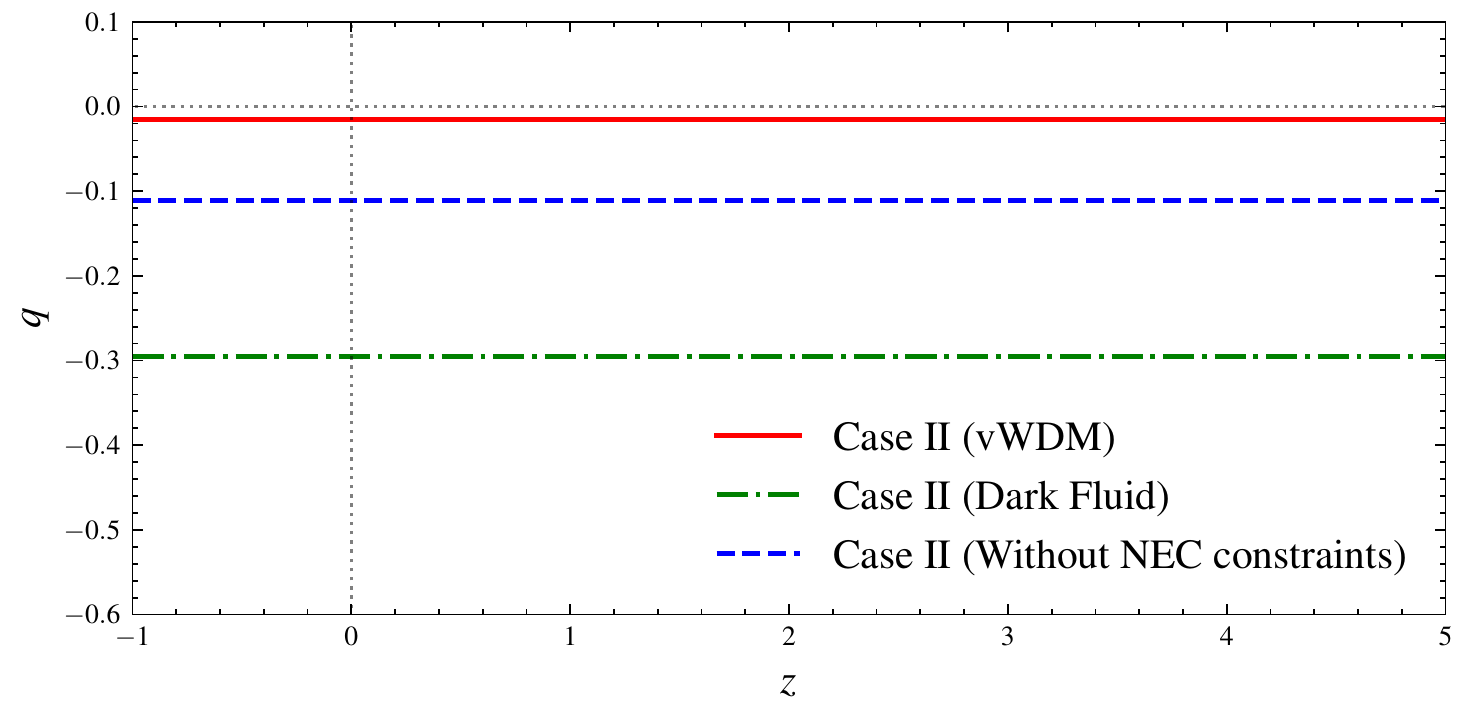}
	\caption{ Plot of deceleration parameter vs redshift corresponding to case II for the best fit parameter values given in table \ref{tab1}.}
	\label{DEC1}
\end{figure}
\begin{figure}
	\centering
	\includegraphics[width=\columnwidth]{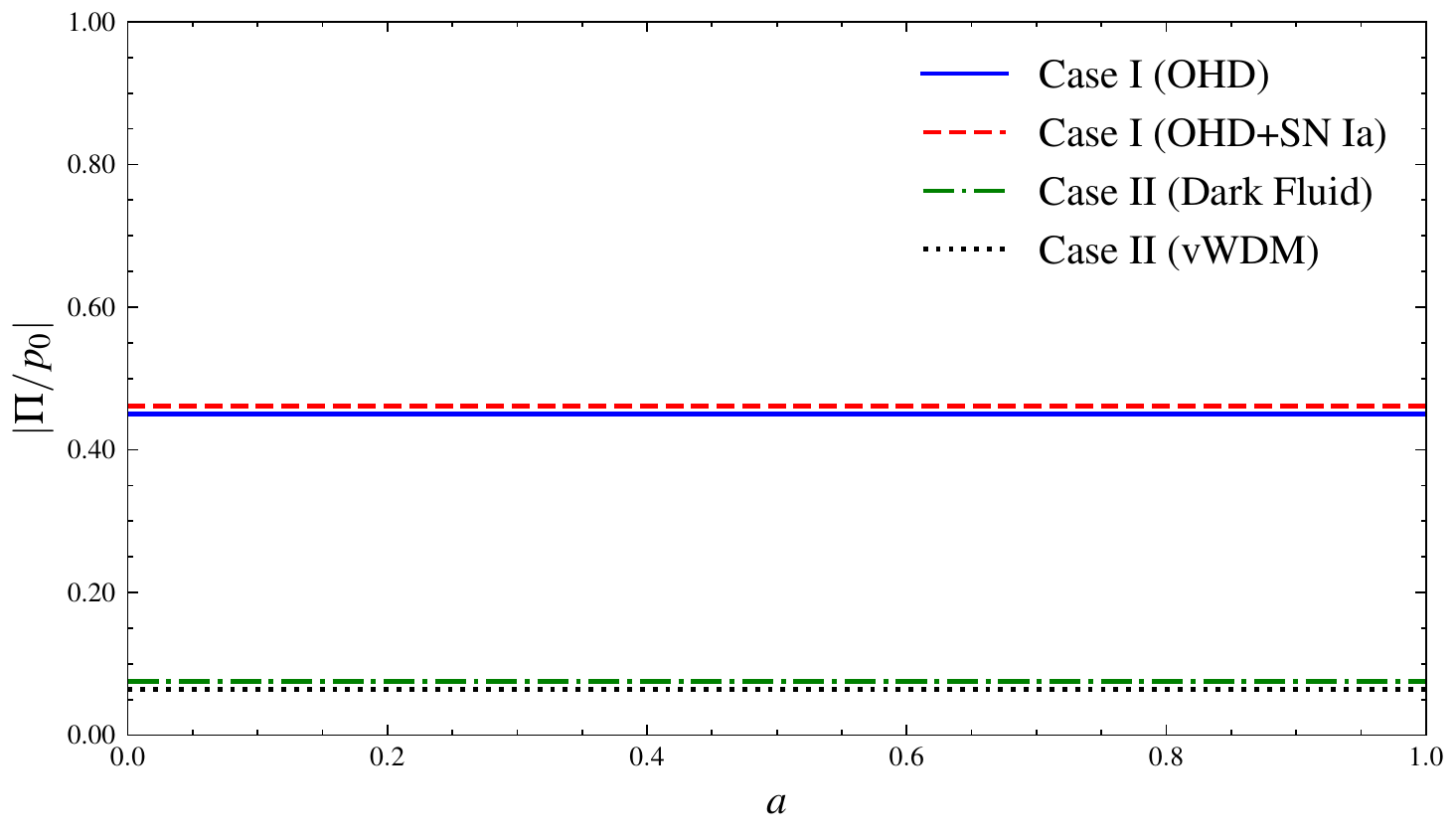}
	\caption{ Plot of near equilibrium condition for both models corresponding to the best fit values of model parameters.}
	\label{NEC1}
\end{figure}
\begin{figure*}
	\centering
	\includegraphics[width=2\columnwidth]{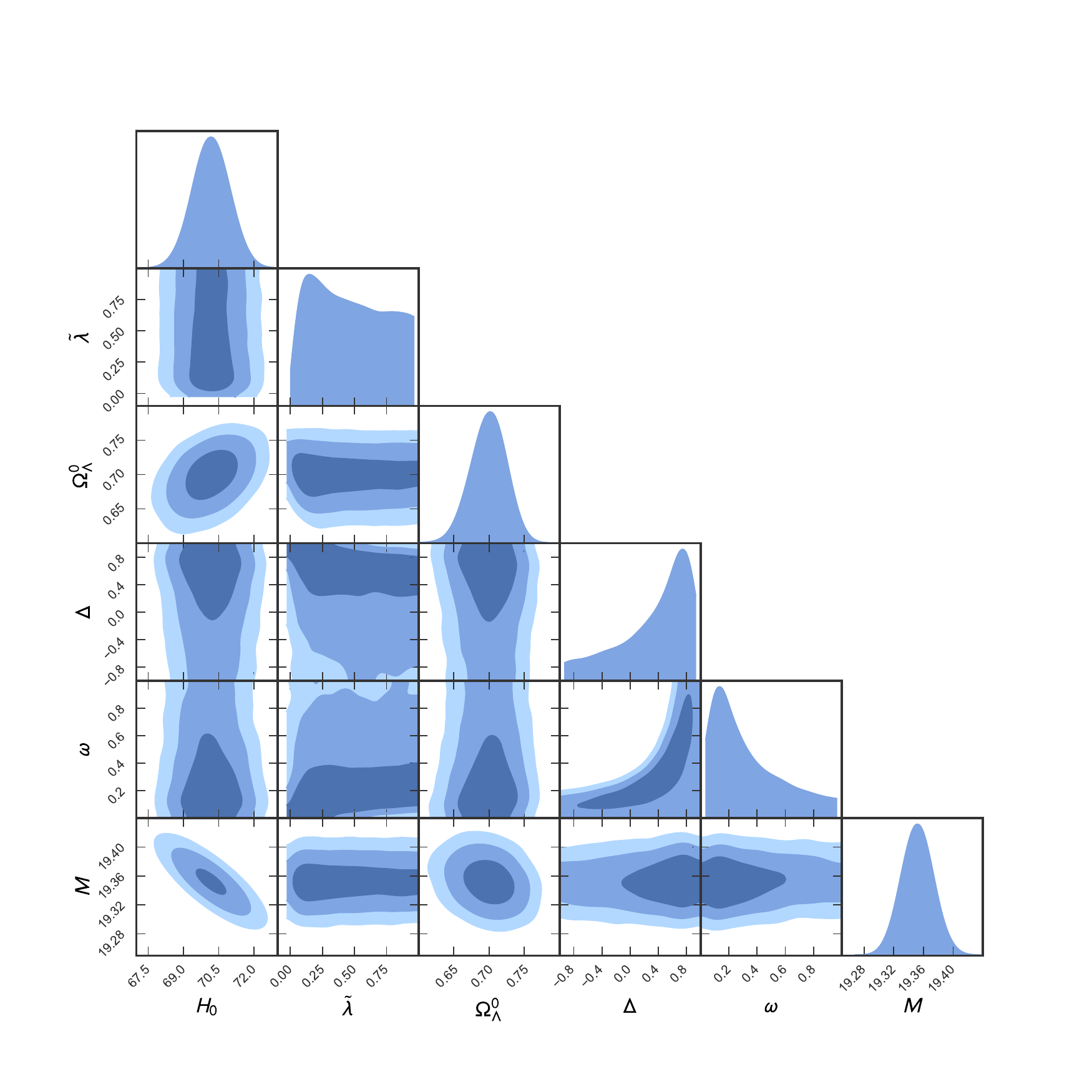}
	\caption{Corner plot of model parameters, obtained from parameter estimation through $\chi^{2}$ minimization using combined OHD+SNe Ia data for case I.
	}
	\label{Contour}
\end{figure*}
For data analysis of case II, we consider three different scenarios. First: the generic fluid shows dark fluid like nature and obeys (\ref{24}), (\ref{29}), (\ref{31})  and (\ref{12}) constraints with prior (\ref{dark}). Second: the generic fluid shows vWDM fluid nature and follows the constraints (\ref{18}), (\ref{18.1}), (\ref{30}), (\ref{32}), (\ref{12}) with prior (\ref{WDM}). And third: we neglect the NEC, and allow the parameters to vary freely, hence assuming only uniform priors (\ref{prior1}) to check the feasibility of the model without the constraints.
\begin{eqnarray} \label{dark}
	\tilde{\lambda}\in [-3/8,1]\;:\; \omega <0 \;:\; H_{0} \in [50,80] 
\end{eqnarray}
\begin{eqnarray} \label{WDM}
	\tilde{\lambda}\in [-1,-3/8]\;:\; \omega >0 \;:\; H_{0} \in [50,80] 
\end{eqnarray}
\begin{eqnarray} \label{prior1}
	\tilde{\lambda}\in [-1,1]\;:\; \omega >0 \;:\;  \zeta_{0}>0\;:\; H_{0} \in [50,80] 
\end{eqnarray}
From analysis of Case II, using OHD data, it is seen that, it attains a $\chi^{2}_{dof}=2.96$ when the generic fluid has a dark fluid nature, $\chi^{2}_{dof}=1.94$ when it has vWDM like nature and $\chi^{2}_{dof}=2.72$ when it has vWDM nature, but the NEC constraints are neglected and parameters are varied freely. Even though case II ( model without $\Lambda$) shows no good fit to the data, it is clear from FIG. \ref{DEC1} and  FIG. \ref{NEC1}, that this model is capable of showing an accelerated expansion that simultaneously satisfies the NEC for the fluid. From the evolution of deceleration parameter with redshift, it is learned that this model describes an ever accelerating universe with no transition phase. Hence, it deviates significantly from current concordance model and is therefore omitted from further analysis. As for the model having $\Lambda$ (i.e. case I), in addition to satisfying NEC, as seen from FIG. $\ref{NEC1}$, good fit results are obtained as shown in table \ref{tab1} with a $\chi^{2}_{dof}=0.74$.

\subsubsection{\textbf{OHD+SNe Ia data set:}}
Since, only the model having $\Lambda$ i.e. case I, shows a good fit to data while simultaneously satisfying the required conditions, it is adopted for a detailed analysis using the combined OHD+SNe Ia data. Corner plot of 2D posterior contours with 1 sigma (68\%), two sigma (95\%) and three sigma confidence level (99.7\%) and 1D marginalized posterior distributions of model parameters for the combined OHD+SNe Ia data plotted using \cite{Bocquet2016} is given in FIG. \ref{Contour}. Along with that, the best estimates of all model parameters for this model (case I) is given in table \ref{tab1}. Using these best fit values of model parameters, evolution of some cosmological parameters associated with this model are studied.
\begin{figure}
	\centering
	\includegraphics[width=\columnwidth]{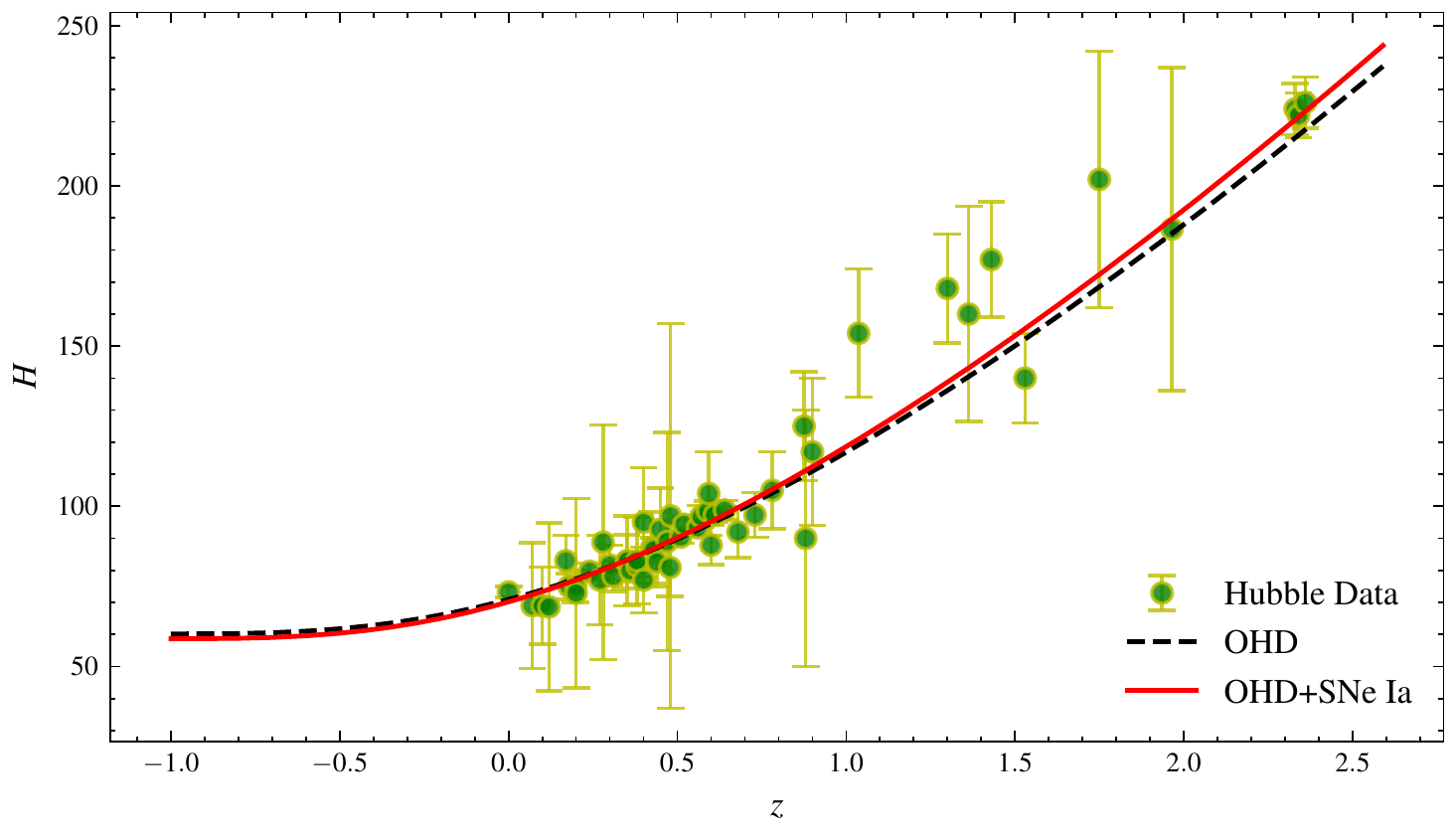}
	\caption{ Graph comparing the evolution of Hubble parameter model associated with case I with best estimated values of model parameters from OHD and combined OHD+SNe Ia datasets, with 52 Hubble data points.}
	\label{OHD1}
\end{figure}
\begin{figure}
	\centering
	\includegraphics[width=\columnwidth]{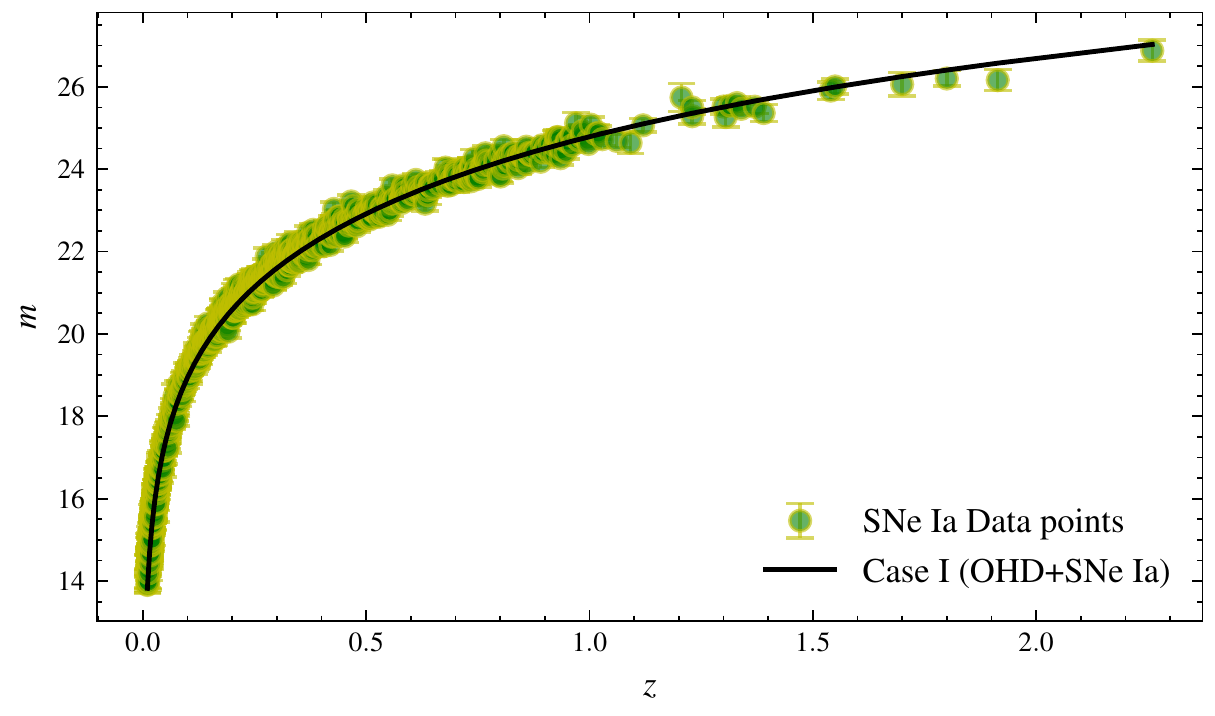}
	\caption{ Graph comparing the evolution of theoretical apparent magnitude for case I with 1048 SNe Ia data points for combined OHD+SNe Ia data within a redshift range of $0.01 \leq z \leq 2.26$.}
	\label{SUP}
\end{figure}
Evolution of the Hubble parameter with redshift, using both the OHD and combined OHD+SNe Ia data is provided in FIG. \ref{OHD1}. The figure confirms that the model predicts an end de Sitter epoch for the evolving universe. Also, the graph comparing theoretical prediction of the apparent magnitude with observed SNe Ia data for case I is also provided as FIG. \ref{SUP}, which shows a reasonable fit of the model with the observed data.

 \paragraph*{\textbf{The near equilibrium condition:}} The plot of the NEC for case I, is shown in FIG. \ref{NEC1}. Here, we see that NEC associated with such a choice of viscous model ($\zeta=\zeta_0 H^{-1}\rho$) remains as a constant value throughout the evolution of the universe. Hence the viscous fluid is always in a near equilibrium state. From this it is clear that, in contrast to both OHD as well as OHD+SNe Ia data, model in case I, obeys the NEC throughout the expansion of the universe. In fact the NEC value is a constant throughout the evolution similar to case II.

\paragraph*{\textbf{Age of the universe:}} Age of the universe predicted by this model can be determined by substituting (\ref{49}) in (\ref{57.1}) and integrating by taking best fit values of all the model parameters. The predicted the age of the universe is then found to be approximately 14.1 Gyr, which is slightly greater than that obtained for $\Lambda$CDM model (13.8 Gyr) \cite{weinberg1972gravitation}. Nevertheless, it is a significant improvement over age predicted by some conventional bulk viscous models \cite{sasidharan2015bulk, mohan2017bulk, nair2016bulk}. 

\paragraph*{\textbf{Coefficient of bulk viscosity:}} The best estimated value of coefficient of bulk viscosity $\zeta$ at present, for this model can be determined from $\zeta=\zeta_{0} H^{-1} \rho$. Substituting this form of viscosity in (\ref{1.8}) and re-arranging, we get the expression for fluid density as,
\begin{equation}\label{60}
	\rho = \frac{3H^2-\Lambda}{\left(1+\tilde{\lambda}\left(3-\omega+3\zeta_{0}\right)\right)}
\end{equation}
The present value of fluid density can be obtained by replacing $H$ by $H_0$ in the above expression. This gives,
\begin{equation}
	\rho^0 = 3H^2_0 \left[\frac{1-\Omega^0_{\Lambda}}{\left(1+\tilde{\lambda}\left(3-\omega+3\zeta_{0}\right)\right)}\right].
\end{equation}
From this, we can determine present value of coefficient of bulk viscosity as,
\begin{equation}
	\zeta^0 = 3\zeta_{0}H_0 \left[\frac{1-\Omega^0_{\Lambda}}{\left(1+\tilde{\lambda}\left(3-\omega+3\zeta_{0}\right)\right)}\right].
\end{equation}
By substituting best fit values of model parameters, we find that value of $\zeta^0$ is equal to $1.95 \times 10^{6} \text{ Pa}\cdot \text{s}$ (in SI units). This value leans towards the lower side of the range of values ($10^{6} - 10^{7} \text{ Pa}\cdot\text{s}$) predicted in some literatures \cite{normann2016general, brevik2015viscosity, brevik2016temperature, sasidharan2016phase}.  
\paragraph*{\textbf{Deceleration parameter:}} For determining the evolution of  deceleration parameter for the model, we use (\ref{63}) and (\ref{49}) along with $\ddot{a}/a$ from (\ref{37}).
\begin{equation}
	\frac{\ddot{a}}{a} = \frac{\left[ \left(1+(3+8\tilde{\lambda})(\omega-3\zeta_{0})\right)\rho - 2 \Lambda \right]}{-6}\label{64}
\end{equation}
Now substituting (\ref{64}) in (\ref{63}) by considering $\rho$ from (\ref{60}) we get the deceleration parameter in terms of scaled Hubble parameter $h=H/H_0$ as,
\begin{multline}\label{65}
	q=\frac{1}{2h^{2}}\left\{   \left(1+(3+8\tilde{\lambda})(\omega-3\zeta_{0})\right) \right.\\
	 \left. \left[ \frac{h^2-\Omega_{\Lambda}}{\left(1+\tilde{\lambda}\left(3-\omega+3\zeta_{0}\right)\right)}\right] - 2 \Omega_{\Lambda}\right\}.
\end{multline}
\begin{figure}
	\centering
	\includegraphics[width=\columnwidth]{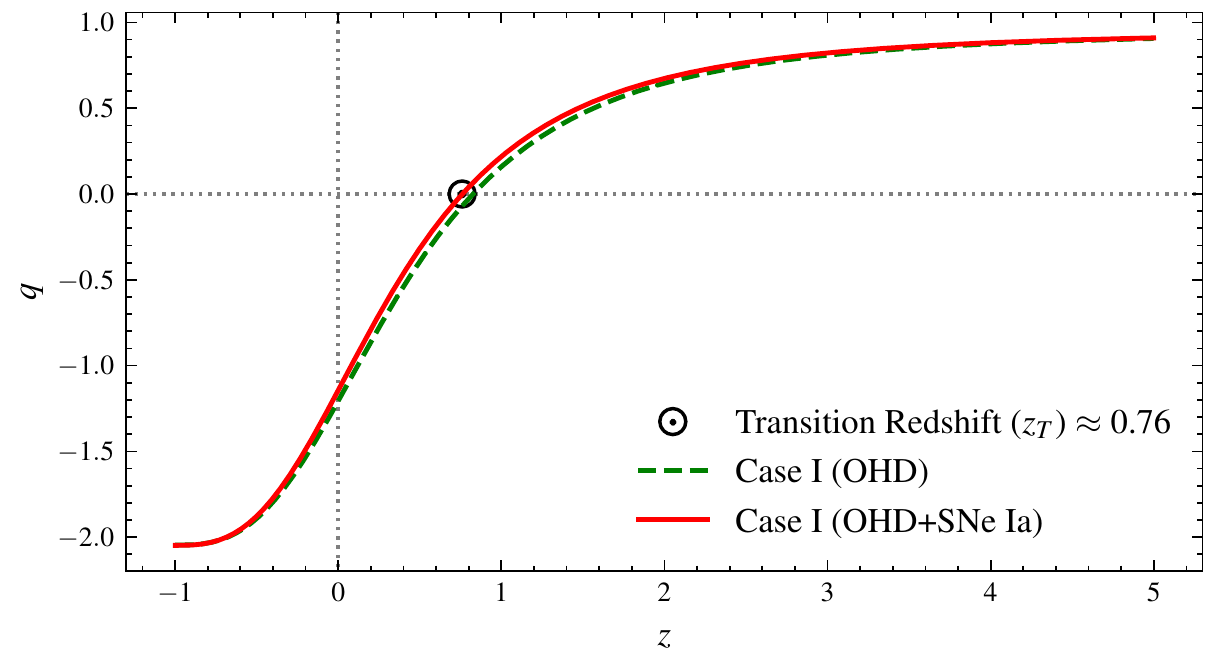}
	\caption{ Plot of deceleration parameter vs redshift for model case I under OHD and combined OHD+SNe Ia analysis.}
	\label{DEC2}
\end{figure}
The evolution of the $q$-parameter with redshift, in this model is presented in FIG. \ref{DEC2}. From analysis of the graph, we can infer that the universe underwent a transition into an accelerated epoch at around a redshift of $z_{T}=0.76.$ This is slightly above than that predicted by the concordance model which is around $z_{T}=0.66.$ The asymptotic behaviour of the deceleration parameter, $q \to -1$ as $z \to -1,$ confirms our previous conclusion that the model predicts an end de Sitter like behavior in the far future of this evolution similar to prediction made by $\Lambda$CDM model.

\section{Discussion and Conclusion}

In this work we have investigated the possibilities of achieving an accelerated expansion of the universe by satisfying NEC for a generic viscous fluid in $f(R,T)$ gravity. We have chosen the functional form of $f(R,T)$ as $R+2\lambda T,$ and carried out the study both in the presence and in the absence of cosmological constant. We first formulate the general constraints for the free parameters of the model, subjected to NEC. These constraints were formulated without assuming a phenomenological form for the viscous pressure $\Pi.$ For developing the constraints, we have assumed a generic viscous fluid as the cosmic component, which can take either negative and positive values for  coefficient of bulk viscosity. The equation of state parameter $\omega$ associated with the kinetic pressure of the same fluid is also assumed to take both negative (Dark Fluid) or positive (WDM) values, but not zero. These assumptions guarantees a most general form for the constraints.

For the model without the cosmological constant, we found the following regarding the constraints. If the generic viscous fluid behaves like vWDM, i.e. with $\omega >0,$ the resulting constraints imposed by NEC, is as given by $\tilde{\lambda}<-3/8$, $\Omega_{\rho}>0$ along with that given in equations (\ref{18}), (\ref{18.1}), (\ref{30}), (\ref{32})  and (\ref{12}). Similarly, if the generic fluid shows dark fluid like behavior with $\omega<0$, NEC is satisfied for $\tilde{\lambda}>-3/8$, $\Omega_{\rho}>0$ along with (\ref{24}), (\ref{29}), (\ref{31}) and (\ref{12}).While for the model with cosmological constant, i.e. with $\Lambda+R+2\lambda T,$ the model parameters are constrained, not by using $\tilde{\lambda},$ but instead by using the coefficient of viscosity and the cosmological constant as seen from (\ref{40}). 

In the next step, we assumed a special case of a newly proposed form for the coefficient of bulk viscosity given by, $\zeta=\zeta_{0}H^{1-2s}\rho^s$. We then solved the Friedmann equations associated with both models (i.e with and without $\Lambda$) by assuming $s=1$ case and then obtained, analytical solutions for Hubble parameter in terms of scale factor which are given in (\ref{49}) for case I and (\ref{43}) for case II. From analysis of these solutions it is seen that, these models are significantly different from each other. We then pointed out a significant drawback of case II, which was that, it is incapable of predicting, a transition from prior deceleration to late acceleration. Then model analysis was done with the aim of assessing the validity of NEC during different epochs of the evolution by imposing the developed constraints on viscous models. 

For the case without cosmological constant, i.e. model case II, we studied three sets of constraints. Firstly, a universe where generic fluid shows viscous dark fluid traits with constant EoS parameter $\omega<0$ and used the constraint set  (\ref{24}), (\ref{29}), (\ref{31}) and (\ref{12}) for applying the expression bounds. Secondly, a universe where generic fluid behaves like vWDM with constant EoS parameter $\omega>0$ and used the constraint set (\ref{18.1}), (\ref{32}) and (\ref{12}). Finally, a universe where generic fluid shows vWDM behavior with $\omega>0$ and using the uniform prior (\ref{prior1}), hence assuming a possible violation of NEC. From analysis of values in table \ref{tab1}, we see that this model is bad fit to data having a minimum $\chi^{2}_{dof}$ of only $1.94$ and is hence neglected from further analysis. Nonetheless, from FIG. \ref{NEC1} and FIG. \ref{DEC1} we learn that, it obeys the NEC throughout the accelerated expansion of the universe. Hence, we make the most important conclusion from this study that, in the context of $R+2\lambda T$ gravity, viscous WDM fluids are able to maintain a near equilibrium state during the accelerated expansion of the universe without the inclusion of cosmological constant provided necessary constraints are obeyed. The bad fit resulted in the analysis of this model is possibly due to poor choice of viscous model and not because of the constraints themselves. Hence in near future, we plan on investigating further viscous models using these developed constraints, since they do not depend on the model that is chosen for bulk viscosity.

Now, for model with the cosmological constant, i.e. case I, we have studied only a single case where, the universe is modeled using a generic fluid behaving like vWDM with constant EoS parameter $\omega>0$ and a positive coefficient of bulk viscosity. Here the NEC is satisfied not by constraining the value of $\tilde{\lambda}$, but by restricting the value of coefficient of bulk viscosity $\zeta$ using relation (\ref{40}). Furthermore, we can infer from FIG. \ref{NEC1}, \ref{OHD1}, \ref{SUP} and table \ref{tab1} we learned that, this model not only shows a good fit to OHD and combined OHD+SNe Ia data, but also satisfies NEC for the viscous fluid throughout the evolution of the universe. Then, using the best estimated values of model parameters, we found that this model predicts the following values for cosmological parameters, which are in the acceptable range based on current literatures. The age of the universe is obtained as, $(t_{0}-t_{b})=14.1$ Gyr. The transition redshift is found to be around $(z_{T})=0.76.$ The Coefficient of bulk viscosity is obtained around $(\zeta) = 1.95 \times 10^{6} \text{ Pa}\cdot \text{s}.$

Form these analyses, we make the following conclusions from this study. Firstly, in the absence of a cosmological constant, the model implies an ever acceleration, however the NEC for the viscous fluid will be satisfied. What one utmost say is that, this result is an aftereffect of the particular form of the viscosity that has been chosen. Secondly, introducing cosmological constant into to the model reduces the number of constraints on model parameters that are needed for satisfying the NEC, i.e presence of cosmological alleviates strictness of NEC constraints. More over the model implies a late accelerated expansion. Thirdly, irrespective of the presence of cosmological constant, the ratio corresponding to the NEC criterion, will be a constant, ie. $\Pi/p_0 =3\zeta_0/w$. Finally, since $\Omega_{\Pi}$ and $\Omega_{\rho}$ depends on exact evolution of the Hubble parameter model, it is important to check the validity of NEC in different regimes of evolution, separately for each bulk viscous model that is considered. This means that the constraints developed are necessary but not sufficient conditions for ensuring NEC for the viscous fluid.

%\bibliographystyle{apsrev}
%\bibliography{ref}

\end{document}